%% file: paper2.tex
\documentclass[%
 aip,
jcp,%
 amsmath,amssymb,
 reprint,%
 groupedaddress,
]{revtex4-1}
\usepackage{graphicx}
\usepackage{dcolumn}
\usepackage{bm}
\usepackage{xcolor}
\usepackage{siunitx}
\usepackage{epstopdf}

\newcommand{\tr}{\mathrm{tr}}
\newcommand{\rd}{\mathrm{d}}

\newcommand{\eu}[1]{\mathrm{e}^{#1}}

\usepackage{soul} 

\usepackage{braket}

\usepackage{dcolumn} 
\newcolumntype{d}[1]{D{.}{.}{#1}} 

\begin{document}

\title{A partially linearized spin-mapping approach for nonadiabatic dynamics.
\\ II. Analysis and comparison with related approaches
}

\author{Jonathan R. Mannouch}
\email{jonathan.mannouch@phys.chem.ethz.ch}
\author{Jeremy O. Richardson}%
\email{jeremy.richardson@phys.chem.ethz.ch}
\affiliation{Laboratory of Physical Chemistry, ETH Z\"{u}rich, 8093 Z\"{u}rich, Switzerland}

\date{\today}

\begin{abstract}
In the previous paper [J. R. Mannouch and J. O. Richardson, J.~Chem.~Phys.~xxx, xxxxx (xxxx)] we derived a new partially linearized mapping-based classical-trajectory technique, called spin-PLDM\@. This method describes the dynamics associated with the forward and backward electronic path integrals, using a Stratonovich--Weyl approach within the spin-mapping space. While this is the first example of a partially linearized spin mapping method, fully linearized spin mapping is already known to be capable of reproducing dynamical observables for a range of nonadiabatic model systems reasonably accurately. Here we present a thorough comparison of the terms in the underlying expressions for the real-time quantum correlation functions for spin-PLDM and fully linearized spin mapping in order to ascertain the relative accuracy of the two methods. In particular, we show that spin-PLDM contains an additional term within the definition of its real-time correlation function, which diminishes many of the known errors that are ubiquitous for fully linearized approaches. One advantage of partially linearized methods over their fully linearized counterparts is that the results can be systematically improved by re-sampling the mapping variables at intermediate times. We derive such a scheme for spin-PLDM and show that for systems for which the approximation of classical nuclei is valid, numerically exact results can be obtained using only a few `jumps'. Additionally, we implement focused initial conditions for the spin-PLDM method, which reduces the number of classical trajectories that are needed in order to reach convergence of dynamical quantities, with seemingly little difference to the accuracy of the result.
\end{abstract}

\maketitle

\section{\label{sec:intro}Introduction}
Trajectory simulations offer a computationally cheap, as well as a physically motivated approach for calculating dynamical quantum-mechanical observables in condensed-phase systems.\cite{Stock2005nonadiabatic} This explains the current popularity of methods such as Ehrenfest dynamics\cite{Ehrenfest1927,Mclachlan1964} and Tully's fewest-switches surface hopping.\cite{Tully1990hopping}

Certain methods within this class can be rigorously derived from a path-integral formulation of real-time quantum correlation functions. Upon linearizing the difference between the forward and backward nuclear paths, the resulting expression for such quantities is ideally suited for evaluation with classical trajectories. In order to simulate nonadiabatic processes, a quantum-classical approach is needed, by which the electronic dynamics can also be described within the same classical-trajectory picture.

One such category of quantum-classical approaches that fits into this formalism are the `fully linearized' methods because the same linearization approximation used for the nuclear paths is also applied to the electronic paths. Examples of such methods are the linearized semiclassical initial-value representation (LSC-IVR)\cite{Miller2001SCIVR,Sun1998mapping,*Wang1999mapping,Shi2004goldenrule,Liu2020} and the Poisson-bracket mapping equation (PBME)\cite{Kim2008Liouville,Kelly2012mapping} approaches, which both make use of the Meyer--Miller--Stock--Thoss (MMST) mapping\cite{Meyer1979spinmatrix,Stock1997mapping} to describe the electronic degrees of freedom, as well as fully linearized spin-mapping,\cite{spinmap,multispin} which uses a Stratonovich--Weyl approach for describing the electronic dynamics within the spin-mapping space. Other forms of mapping also exist, from which fully-linearized methods can be derived.\cite{Liu2016,xin2019,Meyer1979spinmatrix,Cotton2015spin} The spin-mapping formalism appears to result in a more accurate fully linearized method than the MMST formalism; for example fully linearized spin-mapping almost always outperforms LSC-IVR and PBME.\cite{spinmap,multispin} However, we note that there are alternative ways of improving fully linearized MMST mapping-based techniques which make them comparable in accuracy to their spin-mapping analogues.\cite{identity,FMO,Cotton2013SQC,Cotton2013mapping,Cotton2014ET,Cotton2015spin,Miller2015SQC,Miller2016Faraday,Cotton2016_2,Cotton2016SQC,Cotton2017mapping,Liang2018,Cotton2019,Cotton2019SQC} 

Another category of quantum-classical approaches also exists, which encompasses the `partially linearized' methods,\cite{Sun1997} so called because the electronic forward and backward paths, unlike the nuclear paths, are explicitly described by separate dynamical variables. An exact solution of the quantum-classical Liouville equation (QCLE) in terms of independent trajectories does not exist\cite{Kelly2012mapping,Kapral1999} and hence such methods necessarily provide only an approximate solution for the full dynamics of electrons coupled to classical nuclei. Most notable of these are the partially linearized density matrix (PLDM)\cite{Huo2012PLDM,Huo2010,*Huo2011densitymatrix,*Huo2012_2,*Huo2012MolPhys,*Huo2013PLDM,*Huo2015PLDM,Lee2016,Castellanos2017,Mandal2018quasidiabatic,Mandal2018_2,Mandal2019} and the forward-backward trajectory solution (FBTS)\cite{Hsieh2012FBTS,Hsieh2013FBTS,Kapral2015QCL,Kelly2020} approaches, where the electronic dynamics of the two paths are described using MMST mapping.\cite{Meyer1979spinmatrix,Stock1997mapping} In Paper I,\cite{paper1} we developed a new partially linearized method, called spin-PLDM, derived using the Stratonovich--Weyl transform to describe the electronic dynamics within the spin-mapping space. Based on results presented in that work, spin-PLDM appears to reproduce real-time correlation functions with a greater accuracy than other partially linearized methods.

Fully linearized and partially linearized quantum-classical methods are derived using quite different approximations to the underlying path integrals. It is therefore not \emph{a priori} obvious which one of these classes of mapping-based classical-trajectory techniques will consistently produce the most accurate results, although results in the literature suggest that PLDM and FBTS typically outperform PBME and LSC-IVR.\cite{Rekik2013,Kelly2016master,Hoffmann2019,Braver2020} Hence one focus of the current paper is to compare the terms entering the real-time correlation functions for both spin-PLDM and fully linearized spin mapping, in order to further understand the accuracy of computing dynamical observables with both methods. In particular, we show in the proceeding analysis that spin-PLDM incorporates a term with no analogue in the linearized method, which is contained within the definition of the real-time correlation functions of traceless operators. Including this term appears to diminish many of the known errors that are ubiquitous for fully linearized methods.

A clear advantage of partially linearized methods over their fully linearized counterparts is that the accuracy of calculated dynamical observables can be systematically improved to QCLE results by re-sampling the mapping variables at intermediate times, commonly referred to as employing `jumps'. Such schemes have already been developed for partially linearized methods within the MMST mapping space.\cite{Hsieh2013FBTS,Huo2012PLDM} While computationally expensive to convergence, such methods at least enable the error associated with a calculated dynamical observable to be quantified, which is to our knowledge not possible with fully linearized techniques. In this paper we obtain a jump spin-PLDM method and show that typically only a few `jumps' are required to correct the spin-PLDM results towards the QCLE solution, due to the improved accuracy of spin-PLDM over other partially linearized methods. 

One factor that often limits the efficiency and applicability of such mapping-based classical-trajectory methods is the issue of sampling. Often a large number of trajectories are required in order to converge results for dynamical quantities of interest, in particular for systems which contain a large number of degrees of freedom. 
In previous work, focused initial conditions have been developed for mapping-based techniques to alleviate this problem, by restricting the sampling space of initial electronic mapping variables to the region which corresponds to the population of a single initial electronic state. These initial conditions have previously been successfully implemented for fully linearized MMST mapping,\cite{Stock1999ZPE,Kim2008Liouville,Cotton2013SQC,Cotton2013mapping,Cotton2019} fully linearized spin-mapping\cite{spinmap,multispin} and partially linearized MMST mapping techniques,\cite{Bonella2003mapping,Bonella2005LANDmap,Huo2012PLDM,Hsieh2013FBTS} and typically require an order of magnitude fewer trajectories in order to reach the same level of convergence. We show in this paper that similar focused initial conditions can be rigorously derived and also easily implemented for spin-PLDM\@. 
Importantly, the accuracy of the spin-PLDM results is practically unaffected by this modification. We hence believe that spin-PLDM is one of the most promising methods for accurately and efficiently calculating dynamical quantities within condensed-phase nonadiabatic systems. 
\section{Spin Mapping}
\subsection{Fully linearized spin mapping}
Fully linearized spin mapping\cite{spinmap,multispin} describes the electronic dynamics by evolving a single set of classical variables that are constrained to the surface of a hypersphere, where the possible radii of the hypersphere are determined by the Stratonovich--Weyl transform. Within this method, the accuracy of any obtained dynamical quantities, such as real-time correlation functions, depends strongly on the spin-sphere radius that is used. In previous work, it has been found that the so-called W-sphere consistently produces the most accurate results for a wide range of model systems.\cite{spinmap,multispin} This sphere is also the most symmetric choice in the sense that it is `self-dual'.

It is simplest to analyze the method and perform the calculations using the Cartesian representation for the spin variables. In terms of Cartesian mapping variables, this fully linearized W-sphere method, which we will refer to as spin-LSC, has the following expression for the real-time correlation function:
\begin{equation}
\label{eq:corr_LSC}
C_{AB}(t)=\Braket{A_{\text{W}}(\mathcal{Z})B_{\text{W}}(\mathcal{Z}(t))}_{\text{spin-LSC}} ,
\end{equation}
where the $t=0$ value is implied for any quantity for which time, $t$, is not explicitly stated. In this expression, $\mathcal{Z}=\{Z_{1},Z_{2},\cdots,Z_{F}\}$ are the Cartesian mapping variables for an $F$-level system, where $Z_{\lambda}=X_{\lambda}+iP_{\lambda}$ are complex numbers associated with electronic state $\lambda$. These mapping variables represent the electronic degrees of freedom of the system. Along with the nuclear phase-space variables $x$ and $p$, these quantities are initially sampled in the definition of the real-time correlation function from the spin-LSC average, defined as:
\begin{equation}
\label{eq:spin-LSC-av}
\Braket{\cdots}_{\text{spin-LSC}}=\int\rd x\,\rd p\,\rd\mathcal{Z}\cdots\rho_{\text{W}}(\mathcal{Z})\rho_{\text{b}}(x,p) ,
\end{equation}
where $\rd\mathcal{Z}=\prod_{\lambda}\rd X_{\lambda}\rd P_{\lambda}$ and
 $\rho_{\text{b}}(x,p)$ is the Wigner transform of the initial nuclear density matrix, normalized such that $\int\rd x\,\rd p\,\rho_{\text{b}}(x,p)=1$. Additionally, $\rho_{\text{W}}(\mathcal{Z})$ is the initial distribution for the electronic Cartesian mapping variables, which constrains them to lie on the surface of the W-sphere:
\begin{equation}
\label{eq:hypersphere}
\rho_{\text{W}}(\mathcal{Z}) = F\frac{\delta(|\mathcal{Z}|^{2}-R_{\text{W}}^{2})}{\int\rd\mathcal{Z}\,\delta(|\mathcal{Z}|^{2}-R_{\text{W}}^{2})},
\end{equation}
where the factor of $F$ 
ensures the correct normalization such that
$\Braket{1}_\text{spin-LSC}=\text{tr}[\hat{\mathcal{I}}]=F$ and $\text{tr}[\cdots]$ is the partial trace over the electronic degrees of freedom. 
Finally, $R_{\text{W}}$ is the radius of the W-sphere:
\begin{equation}
\label{eq:wsphere_radius}
R_{\text{W}}^{2}=2\sqrt{F+1} .
\end{equation}
Because the Cartesian mapping variables are sampled uniformly from the hypersphere in Eq.~(\ref{eq:spin-LSC-av}), we refer to this as full-sphere initial conditions for the electronic degrees of freedom within the spin-LSC technique.

For the spin-LSC correlation function, given in Eq.~(\ref{eq:corr_LSC}), the electronic operators $\hat{A}$ and $\hat{B}$ are represented by their Stratonovich--Weyl W-functions. The spin-LSC method can also be applied to real-time correlation functions where $\hat{A}$ and $\hat{B}$ contain nuclear operators, although we will not consider such correlation functions in this paper. The W-function of operator $\hat{A}$ is:
\begin{equation}
\label{eq:w_func}
A_{\text{W}}(\mathcal{Z})=\text{tr}[\hat{A}\hat{w}_{\text{W}}(\mathcal{Z})] ,
\end{equation}
which is defined in terms of the Stratonovich--Weyl kernel, $\hat{w}_{\text{W}}(\mathcal{Z})$, 
whose matrix elements are
\begin{equation}
\label{eq:kernel_cartesian}
\braket{\mu|\hat{w}_{\text{W}}(\mathcal{Z})|\lambda}=\tfrac{1}{2}\left(Z_{\mu}Z^{*}_{\lambda}-\gamma_{\text{W}}\delta_{\lambda\mu}\right) .
\end{equation}
The Stratonovich--Weyl kernel contains a zero-point energy parameter, $\gamma_{\text{W}}$, which for the W-spin sphere is given by:
\begin{equation}
\label{eq:gamma_w}
\gamma_{\text{W}}=\frac{1}{F}(R_{\text{W}}^{2}-2) .
\end{equation}
Inserting Eq.~(\ref{eq:kernel_cartesian}) into  Eq.~(\ref{eq:w_func}) leads to an equivalent expression for the W-function of operator $\hat{A}$:
\begin{equation}
\label{eq:w_func2}
A_{\text{W}}(\mathcal{Z})=\tfrac{1}{2}\sum_{\lambda,\mu}\braket{\lambda|\hat{A}|\mu}\left(Z^{*}_{\lambda}Z_{\mu}-\gamma_{\text{W}}\delta_{\lambda\mu}\right) .
\end{equation}

The spin-LSC correlation function [Eq.~(\ref{eq:corr_LSC})] approximates the $\hat{B}$ operator at time $t$ by evolving in time the Cartesian mapping variables, $\mathcal{Z}$, for the electronic degrees of freedom. These variables, along with the nuclear phase-space variables, are propagated under 
the following
equations of motion:
\begin{equation}
\label{eq:MMST_eom}
\begin{split}
&\frac{\rd Z_{\lambda}}{\rd t}=-i\sum_{\mu}\braket{\lambda|\hat{V}(x)|\mu}Z_{\mu} , \\
&\frac{\rd x}{\rd t}=\frac{p}{m} , \\
&\frac{\rd p}{\rd t}=F_{0}(x)+F_{\text{e}}(\mathcal{Z},x) ,
\end{split}
\end{equation}
which despite the different derivation are completely equivalent to those of the MMST fully linearized mapping approaches and thus recovers the exact electronic dynamics for an isolated subsystem.\cite{spinmap} This is not necessarily the case for other dynamical methods based on spin analogies. \cite{Cotton2015spin}  Within these equations of motion, the potential energy surface as a function of nuclear configuration has been partitioned into two components: $V_{0}(x)+\hat{V}(x)$. The former is a state-independent potential, whereas $\hat{V}(x)$ includes electron-nuclear coupling and is defined such that $\text{tr}[\hat{V}(x)]=0$ for all values of $x$. The corresponding nuclear forces are given by:
\begin{subequations}
\begin{align}
F_{0}(x)&=-\nabla V_{0}(x) , \\
F_{\text{e}}(\mathcal{Z},x)&=-\nabla V_{\text{m}}(\mathcal{Z}) ,
\end{align}
\end{subequations}
where $\nabla$ is the gradient, a vector of derivatives with respect to the nuclear positions. It has been observed from numerical simulations that spin-LSC generally gives rise to more accurate correlation functions than fully linearized approaches based on MMST mapping, in particular for $\hat{A}=\hat{\mathcal{I}}$ in asymmetric systems, where $C_{\mathcal{I}B}(t)$ does not tend to zero in the long-time limit. For these identity-containing correlation functions, it can be shown that within spin-LSC they have the following simple form:  
\begin{equation}
\label{eq:lsc_ident}
C_{\mathcal{I}B}(t)=\Braket{B_{\text{W}}(\mathcal{Z}(t))}_{\text{spin-LSC}} .
\end{equation}
which means that the identity operator is treated exactly (i.e.\ as the number 1) within the spin-LSC method. Whereas this appears naturally in fully linearized spin-mapping methods, a modification of MMST mapping was required to obtain a similar result.\cite{identity,FMO,linearized}
\subsection{\label{sec:spin-PLDM}Spin-PLDM}
Within the spin-LSC method, the Stratonovich--Weyl kernels are used to describe the observable electronic operators, $\hat{A}$ and $\hat{B}$, that appear within the real-time correlation function. The electronic dynamics is thus described by a single set of Cartesian electronic mapping variables, $\mathcal{Z}=\{Z_{1},Z_{2},\cdots,Z_{F}\}$. In contrast, the spin-PLDM method, derived in Paper I,\cite{paper1} is a partially linearized approach, where the Stratonovich--Weyl kernels are used to describe the forward and backward real-time propagators. This leads to an expression for the real-time correlation function which contains two sets of Cartesian mapping variables, $\mathcal{Z}$ and $\mathcal{Z}'$:
\begin{equation}
\label{eq:corr_PLDM}
C_{AB}(t)=\Braket{\tr\left[\hat{A}\hat{w}^{\dagger}_{\text{W}}(\mathcal{Z}',t)\hat{B}\hat{w}_{\text{W}}(\mathcal{Z},t)\right]}_{\text{spin-PLDM}} .
\end{equation}
The initial sampling of the Cartesian mapping variables, $\mathcal{Z}$ and $\mathcal{Z}'$, along with the nuclear phase-space variables, $x$ and $p$, is defined by the spin-PLDM average, given by:
\begin{equation}
\label{eq:spin-pldm-av}
\Braket{\cdots}_{\text{spin-PLDM}}=\int\rd x\,\rd p\,\rd\mathcal{Z}\,\rd\mathcal{Z}'\,\cdots\rho_{\text{W}}(\mathcal{Z})\rho_{\text{W}}(\mathcal{Z}')\rho_{\text{b}}(x,p) .
\end{equation}
Because each set of Cartesian mapping variables in Eq.~(\ref{eq:spin-pldm-av}) is uniformly sampled from the surface of a hypersphere, as for spin-LSC, we refer to this as full-sphere initial conditions for the electronic degrees of freedom within the spin-PLDM technique. 

In the spin-PLDM correlation function, the time-evolved W-kernel, $\hat{w}_{\text{W}}(\mathcal{Z},t)$, can be obtained by applying the time-ordered propagator, $\hat{U}(t)$, to the left of the W-kernel, defined by Eq.~(\ref{eq:kernel_cartesian}):
\begin{equation}
\label{eq:kernel_cartesian_time}
\braket{\mu|\hat{w}_{\text{W}}(\mathcal{Z},t)|\lambda}=\tfrac{1}{2}\left(Z_{\mu}(t)Z^{*}_{\lambda}-\gamma_{\text{W}}\braket{\mu|\hat{U}(t)|\lambda}\right) ,
\end{equation}
where:
\begin{equation}
\label{eq:time-ordered-prop}
\hat{U}(t)=\eu{-i\hat{V}(x(t_{N}))\epsilon}\cdots\eu{-i\hat{V}(x(t_{2}))\epsilon}\,\eu{-i\hat{V}(x(t_{1}))\epsilon} .
\end{equation}
In addition, $t_{k}=kt/N$ is the time at each time-step of the propagation and $N$ is the number of time-steps. As for all classical-trajectory methods, we wish to approach the $N\rightarrow\infty$ limit when performing numerical calculations. The variables are evolved in spin-PLDM under the same equations of motion as Eq.~(\ref{eq:MMST_eom}) with $\mathcal{Z}'$ treated equivalently to $\mathcal{Z}$, except that the expression for the electronic dependent nuclear force is now given by:
\begin{equation}
F_{\text{e}}(\mathcal{Z},\mathcal{Z}',x)=-\tfrac{1}{2}\left[\nabla V_{\text{m}}(\mathcal{Z})+\nabla V_{\text{m}}(\mathcal{Z}')\right] .
\end{equation}
Hence the Cartesian mapping variables for the forward and backward paths, $\mathcal{Z}$ and $\mathcal{Z'}$ respectively, are coupled in the equations of motion via this nuclear force term.

As with all mapping-based classical-trajectory techniques, the efficiency of these spin-mapping methods is often limited by the number of trajectories one requires to converge the results. This is potentially a greater issue for spin-PLDM, because the initial sampling now contains integrals over two sets of Cartesian mapping variables.\footnote{We are currently using uniform sampling over the surface of the sphere. One could of course conceive of improved Monte Carlo schemes which would improve the efficiency without requiring focusing.} However, such convergence issues can be alleviated using focused initial conditions of the initial Cartesian mapping variables, which we will first introduce for spin-LSC.
\section{Focused Sampling}
\subsection{Spin-LSC}\label{sec:foc_lsc}
Focused sampling has been previously implemented for spin-LSC in Refs.~\onlinecite{spinmap} and \onlinecite{multispin}. In this subsection, we present spin-LSC focused initial conditions in a way that makes clear the connection with focused sampling used in spin-PLDM, which will be introduced in Sec.~\ref{sec:foc}. For the case of real-time correlation functions with an initial population operator, our approach becomes identical to these previously implemented focused initial conditions, but slightly different if starting from non-diagonal operators.

The spin-LSC method described above makes use of the properties of the Stratonovich--Weyl kernel. In particular, the full-sphere method relies on the fact that the trace of two operators can be written as an integral over their corresponding W-functions:
\begin{equation}
\label{eq:sw_full}
\text{tr}[\hat{A}\hat{B}]=\int\rd\mathcal{Z}\,\rho_{\text{W}}(\mathcal{Z})A_{\text{W}}(\mathcal{Z})B_{\text{W}}(\mathcal{Z}) , 
\end{equation}
where $\rho_{\text{W}}(\mathcal{Z})$ is the initial distribution for the Cartesian mapping variables corresponding to the surface of the W-sphere [Eq.~(\ref{eq:hypersphere})] and $A_{\text{W}}(\mathcal{Z})$ and $B_{\text{W}}(\mathcal{Z})$ are the W-functions of the corresponding electronic operators [Eq.~(\ref{eq:w_func2})]. 

This property is, however, also satisfied when using alternative sampling distributions of the Cartesian mapping variables, such as focused conditions:
\begin{equation}
\label{eq:sw_foc}
\text{tr}[\hat{A}\hat{B}]=\sum_{\lambda}\int\rd\mathcal{Z}\,\rho^{(\lambda)}_{\text{foc}}(\mathcal{Z})A_{\text{W}}(\mathcal{Z})B_{\text{W}}(\mathcal{Z}) . 
\end{equation}
A proof showing that these focused conditions do satisfy this W-functions property and thus preserve all properties of the Stratonovich-Weyl transform for any electronic operators $\hat{A}$ and $\hat{B}$, is given in Appendix~\ref{app:foc}. These focused conditions are similar in form to the full-sphere sampling of the initial Cartesian mapping variables, given by Eq.~(\ref{eq:sw_full}), but with the following differences. Instead of simply constraining the mapping variables to lie on a hypersphere defined by $|\mathcal{Z}|^{2}=R_{\text{W}}^{2}$, the mapping variables in the focused initial conditions are instead now constrained by $\rho_{\text{foc}}^{(\lambda)}(\mathcal{Z})$, the focused sampling distribution for electronic state $\lambda$:
\begin{equation}
\label{eq:foc_distrib}
\rho_{\text{foc}}^{(\lambda)}(\mathcal{Z})=\frac{\delta\left(|Z_{\lambda}|^{2}-\gamma_{\text{W}}-2\right)\prod_{\mu\neq\lambda}\delta\left(|Z_{\mu}|^{2}-\gamma_{\text{W}}\right)}{\int\rd\mathcal{Z}\,\delta\left(|Z_{\lambda}|^{2}-\gamma_{\text{W}}-2\right)\prod_{\mu\neq\lambda}\delta\left(|Z_{\mu}|^{2}-\gamma_{\text{W}}\right)} .
\end{equation}
This distribution, $\rho_{\text{foc}}^{(\lambda)}(\mathcal{Z})$, still ensures the mapping variables are on the surface of the hypersphere, but also further constrains them so that they entirely occupy electronic state $\ket{\lambda}$ in the sense that:
\begin{equation} \label{eq:pop_foc}
[\ket{\mu}\bra{\mu}]_{\text{W}}(\mathcal{Z})=\delta_{\mu\lambda} ,
\end{equation}
for all $\mathcal{Z}$ with non-zero values of $\rho_\text{foc}^{(\lambda)}(\mathcal{Z})$. Additionally, the factor of $F$ within the definition of the full-sphere initial sampling in Eq.~(\ref{eq:spin-LSC-av}) is now incorporated within the focused variant into the sum over $\lambda$, a complete set of orthonormal electronic states onto which the Cartesian mapping variables are focused.

From the definition of the focused sampling in Eq.~(\ref{eq:sw_foc}), the focused spin-LSC correlation function then becomes:
\begin{equation}
\label{eq:corr_lsc_foc}
C_{AB}(t)=\Braket{A_{\text{W}}(\mathcal{Z})B_{\text{W}}(\mathcal{Z}(t))}_{\text{spin-LSC}}^{\text{foc}} ,
\end{equation}
where the focused spin-LSC average is defined as follows:
\begin{equation}
\label{eq:spin-LSC-foc}
\Braket{\cdots}_{\text{spin-LSC}}^{\text{foc}}=\sum_{\lambda}\int\rd x\,\rd p\,\rd\mathcal{Z}\cdots\rho_{\text{foc}}^{(\lambda)}(\mathcal{Z})\rho_{\text{b}}(x,p) .
\end{equation}
The definition of this real-time correlation function is defined so that the focusing, given by Eq.~(\ref{eq:spin-LSC-foc}) is only performed at $t=0$. This is in contrast to the symmetrical quasi-classical (SQC) windowing method, where windowing functions are used to essentially focus trajectories at time $t$ as well.\cite{Cotton2013SQC,Cotton2013mapping,Cotton2019} For real-time correlation functions with an initial population operator (i.e., $\hat{A}=\ket{n}\bra{n}$), the focused initial sampling defined in Eq.~(\ref{eq:spin-LSC-av}) can be simplified as follows. Using the expression for the W-function of a population operator under focused initial conditions, given by Eq.~(\ref{eq:pop_foc}), the expression for this focused spin-LSC correlation function then becomes:
\begin{equation}
\begin{split}
\langle[\ket{n}\bra{n}]_{\text{W}}&(\mathcal{Z})B_{\text{W}}(\mathcal{Z}(t))\rangle_{\text{spin-LSC}}^{\text{foc}}= \\
&\int\rd x\,\rd p\,\rd\mathcal{Z}\,B_{\text{W}}(\mathcal{Z}(t))\rho_{\text{foc}}^{(n)}(\mathcal{Z})\rho_{\text{b}}(x,p) ,
\end{split}
\end{equation}
which is identical to the expression used in Refs.~\onlinecite{spinmap} and \onlinecite{multispin}. The fact that only one term in the sum over $\lambda$ in Eq.~(\ref{eq:spin-LSC-av}) contributes to the spin-LSC real-time correlation function when $\hat{A}=\ket{n}\bra{n}$ means that focused initial conditions are particularly simple in this case. To treat correlations with off-diagonal (coherence) $\hat{A}$ operators, however, one must include all of the terms in this sum over $\lambda$ in Eq.~(\ref{eq:spin-LSC-foc}). Note that an alternative approach to implement focused initial conditions more efficiently in this case was suggested in Refs.~\onlinecite{spinmap} and \onlinecite{Montoya2016GQME} by choosing a basis which diagonalizes the initial operator $\hat{A}$.

\begin{figure}
\resizebox{0.35\textwidth}{!}{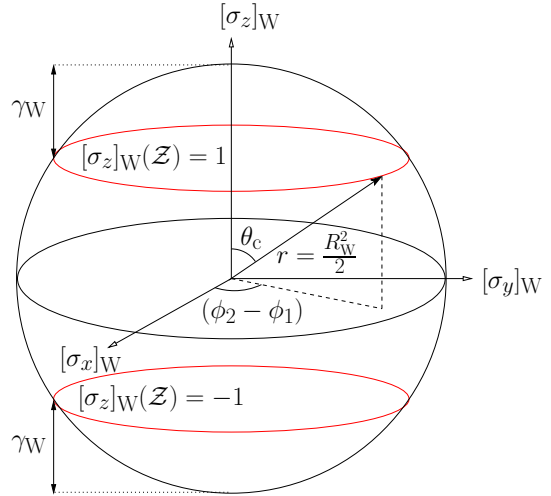}
\caption{An illustration of the two `polar circles' (red) on the W-spin sphere for a two-level system, from which the Cartesian mapping variables are sampled when using focused initial conditions. The upper arctic circle, having a latitude of $\cos\theta_{\text{c}}=1/\sqrt{3}$, corresponds to the electronic system solely occupying state $\ket{1}$ (i.e., $[\sigma_{z}]_{\text{W}}(\mathcal{Z})=1$) and the antarctic circle to state $\ket{2}$.}\label{fig:spin}
\end{figure}

The constraints imposed within focused initial conditions can easily be implemented using the following parameterization of the Cartesian mapping variables:\cite{multispin}
\begin{equation}
\label{eq:arctic_param}
Z_{\mu}=r_{\mu}\eu{i\phi_{\mu}} ,
\end{equation}
where $\phi_{\mu}$ is uniformly sampled from $0$ to $2\pi$ and when focused onto $\lambda$, $r_{\mu}$ is given by:
\begin{equation}
\label{eq:r_def}
r_{\mu}=\sqrt{2\delta_{\mu\lambda}+\gamma_{\text{W}}} .
\end{equation}
In order to illustrate these focused conditions further, we consider the case of a two-level system, in which the electronic state can be described in terms of the expectation values of the Pauli spin matrices and the identity operator. From Eq.~(\ref{eq:arctic_param}), these expectation values for focused initial conditions are:
\begin{subequations}
\begin{align}
& \mathcal{I}_{\text{W}}(\mathcal{Z})=1 , \\
&[\sigma_{x}]_{\text{W}}(\mathcal{Z})=\sqrt{\gamma_{\text{W}}(2+\gamma_{\text{W}})}\,\cos(\phi_{2}-\phi_{1}) , \\
&[\sigma_{y}]_{\text{W}}(\mathcal{Z})=\sqrt{\gamma_{\text{W}}(2+\gamma_{\text{W}})}\,\sin(\phi_{2}-\phi_{1}) , \\
&[\sigma_{z}]_{\text{W}}(\mathcal{Z})=\pm1 ,
\end{align}
\end{subequations}
where the sign of the $\sigma_{z}$ expectation value corresponds to the initial state for which the mapping variables are focused onto. Hence each Cartesian mapping variable is focused onto one of two `polar circles', as illustrated in Fig.~\ref{fig:spin}, and \mbox{$(\phi_{2}-\phi_{1})$} can be thought of as the azimuthal angle around this circle, whereas $(\phi_1+\phi_2)$ is an unimportant cyclic variable. In addition, the upper arctic circle satisfies $[\sigma_{z}]_{\text{W}}(\mathcal{Z})=1$ and the lower antarctic circle satisfies $[\sigma_{z}]_{\text{W}}(\mathcal{Z})=-1$. 

The focused initial conditions are basis-dependent and hence there is a choice of which complete set of states to focus onto. Focusing onto the electronic states which diagonalize the nuclear force operator guarantees that the nuclear dynamics are correct in the absence of off-diagonal diabatic couplings. This is therefore advantageous for systems far from the Born--Oppenheimer limit. Additionally, focusing onto the adiabatic electronic states would be advantageous for systems close to the Born--Oppenheimer limit. Note also that the dynamics can be carried out equivalently in this representation.\cite{Cotton2017mapping} The model systems that we consider in this paper are all far from the Born--Oppenheimer limit, as this is the regime that is traditionally most difficult for mapping-based classical-trajectory techniques to describe correctly. Hence, from now on we will only use diabatic focusing.

For the spin-LSC method, the difference in the converged results between full-sphere and focused sampling was hardly noticeable in our tests, but the latter required an order of magnitude fewer trajectories. This is consistent with the results presented in Refs.~\onlinecite{spinmap} and \onlinecite{multispin}, where the difference between computed correlations functions using full-sphere and focused sampling within the spin-LSC method (referred to in these papers as the W-method) are again hardly noticeable. Note that this is by no means true of other mapping approaches; for example spin-mapping on the Q-sphere appears to be fairly accurate when using full-sphere sampling, but with focusing reduces to Ehrenfest dynamics, which is known to be inaccurate. 
From now on, the numerical results presented in this paper will only be for the focused variant of spin-LSC\@. However, the proceeding analysis and conclusions that we present are equally valid for both full-sphere and focused variants. 
\subsection{\label{sec:foc}Spin-PLDM}
For spin-LSC, the key property required for a sampling scheme 
is that it preserves the properties of the Stratonovich--Weyl kernel in Eq.~(\ref{eq:sw_full}). For spin-PLDM, we require that the focused sampling can correctly describe electronic quantum operators in terms of the W-kernel. This is satisfied by the full-sphere sampling approach:
\begin{equation}
\label{eq:kernel_prop_full}
\hat{A}=\int\rd\mathcal{Z}\,\rho_{\text{W}}(\mathcal{Z})\hat{w}_{\text{W}}(\mathcal{Z})A_{\text{W}}(\mathcal{Z})
\end{equation}
and also by the focused initial conditions: 
\begin{equation}
\label{eq:kernel_prop_foc}
\hat{A}=\sum_{\lambda}\int\rd\mathcal{Z}\,\rho^{(\lambda)}_{\text{foc}}(\mathcal{Z})\hat{w}_{\text{W}}(\mathcal{Z})A_{\text{W}}(\mathcal{Z}) ,
\end{equation}
as can be proved using the same arguments that are presented in Appendix~\ref{app:foc}. In spin-PLDM, two sets of Cartesian mapping variables are used, in order to represent both the forward and backward electronic paths. Hence in focused spin-PLDM, both sets of mapping variables must be initially sampled independently using these focused conditions. This leads to the following expression for the correlation function:
\begin{equation}
\label{eq:corr_pldm_foc}
C_{AB}(t)=\Braket{\tr\left[\hat{A}\hat{w}^{\dagger}_{\text{W}}(\mathcal{Z}',t)\hat{B}\hat{w}_{\text{W}}(\mathcal{Z},t)\right]}_{\text{spin-PLDM}}^{\text{foc}} ,
\end{equation}
where the ensemble average now amounts to using focused conditions for both sets of Cartesian mapping variables:
\begin{equation}
\label{eq:spin-PLDM-foc}
\begin{split}
&\Braket{\cdots}_{\text{spin-PLDM}}^{\text{foc}}= \\
&\sum_{\lambda,\lambda'}\int\rd x\,\rd p\,\rd\mathcal{Z}\,\rd\mathcal{Z}'\cdots\rho_{\text{foc}}^{(\lambda)}(\mathcal{Z})\rho_{\text{foc}}^{(\lambda')}(\mathcal{Z}')\rho_{\text{b}}(x,p) .
\end{split}
\end{equation}
Additionally, the factor of $F^{2}$ within the definition of the spin-PLDM full-sphere sampling, given in Eq.~(\ref{eq:spin-pldm-av}), is incorporated within the focused variant by two independent sums over the complete set of electronic states ($\ket{\lambda}$ and $\ket{\lambda'}$).

In contrast to the focused spin-LSC method, the focused spin-PLDM method contains terms where the initial Cartesian mapping variables for the forward and backward propagator paths are focused onto different electronic states, $\ket{\lambda}$ and $\ket{\lambda'}$. Because fully linearized methods only consider the average of these two paths, such configurations cannot be represented in spin-LSC.
The focused conditions previously implemented for standard PLDM and FBTS\cite{Hsieh2013FBTS} also focus both the forward and backward paths onto the same initial electronic state, which can lead to poor results for long-time dynamics or for systems with relatively strong electron-nuclear coupling, degrading the results relative to standard PLDM/FBTS sampling.\cite{Huo2012PLDM,Hsieh2013FBTS} Within the spin-PLDM approach, however, focusing leads to almost the same result as with full-sphere sampling.  Because the Stratonovich--Weyl kernels are used to represent the forward and backward propagators in spin-PLDM, rather than the observable operators, none of the terms in the $\{\lambda,\lambda'\}$ sum in Eq.~(\ref{eq:spin-PLDM-foc}) is identically zero for the correlation function given in Eq.~(\ref{eq:corr_pldm_foc}), even when $\hat{A}=\ket{n}\bra{n}$. This means that although focused spin-PLDM is a much more efficient method than performing spin-PLDM with full-sphere sampling, it still require more terms to be calculated than for spin-LSC.\footnote{The spin-PLDM focused conditions are certainly less efficient than those for spin-LSC, when these conditions are implemented by uniformly sampling each of the $F^{2}$ terms. However not all of these $F^{2}$ terms in focused spin-PLDM are of equal magnitude and hence more effective sampling schemes could certainly be used to improve the efficiency.} 

As for the spin-LSC focused conditions, the sampling of focused spin-PLDM is now basis-dependent. However, using either a diabatic or adiabatic basis seems the obvious choice, depending on whether the system is close to or far away from the Born--Oppenheimer limit. 
As stated before, we will only use diabatic focusing in this paper.
\section{Jump Spin-PLDM}
One advantage of partially linearized methods over their fully linearized counterparts is that the associated results can be systematically approved towards an exact solution of the quantum-classical Liouville equation. This can be achieved by re-sampling the Cartesian mapping variables at intermediate times, in order to relax the approximations that were made within the derivation of standard PLDM/FBTS and spin-PLDM to describe the electronic transition amplitudes in terms of continuous classical trajectories. Such schemes, called jump FBTS and iterative PLDM, have already been implemented for standard PLDM/FBTS.\cite{Hsieh2013FBTS,Huo2012PLDM} While the scheme is generally inefficient because the computational cost scales exponentially with the number of `jumps', the method can still provide a useful way of quantifying the error associated with dynamical quantities of interest, particularly because the computational cost for performing only a few jumps is relatively cheap.\cite{Kelly2020_2} Following the derivation of jump FBTS, the analogous expression for the real-time correlation function associated with jump spin-PLDM is obtained as: 
\begin{equation}
\begin{split}
\label{eq:corr_jump_PLDM}
C_{AB}(t)=\Bigg\langle\tr&\left[\hat{A}\left(\prod_{m=0}^{n_{\text{jump}}}\hat{w}_{\text{W}}(\mathcal{Z}'_{m},t_{m})\right)^{\dagger}\right. \\
&\left.\qquad\times\hat{B}\prod_{m=0}^{n_{\text{jump}}}\hat{w}_{\text{W}}(\mathcal{Z}_{m},t_{m})\right]\Bigg\rangle^{\text{jump}}_{\text{spin-PLDM}} ,
\end{split}
\end{equation}
where the product, $\prod_{m=0}^{n_{\text{jump}}}$, places terms of successively increasing $m$ to the left of the previous terms. Eq.~(\ref{eq:corr_jump_PLDM}) consists of performing $n_{\text{jump}}$ re-samplings or `jumps' for both sets of Cartesian mapping variables after each time interval $\{t_{0},\cdots,t_{n_{\text{jump}}-1}\}$. The time intervals must be chosen to satisfy $t=\sum_{m=0}^{n_{\text{jump}}}t_{m}$.

Because the sampling of Cartesian mapping variables using either focused or full-sphere conditions allows any electronic operator to be exactly represented by a Stratonovich-Weyl kernel, as illustrated in Eqs.~(\ref{eq:kernel_prop_full}) and (\ref{eq:kernel_prop_foc}), either sampling scheme could be used to perform the `jumps' within jump spin-PLDM. Due to the improved convergence properties associated with focused conditions, this will be used in our jump spin-PLDM scheme. This means that the jump spin-PLDM average is defined as:
\begin{equation}
\label{eq:jump-spin-PLDM}
\begin{split}
&\Braket{\cdots}_{\text{spin-PLDM}}^{\text{jump}}=\int\rd x\,\rd p\,\rho_{\text{b}}(x,p) \\
&\times\int\prod_{m=0}^{n_{\text{jump}}}\rd\mathcal{Z}_{m}\,\rd\mathcal{Z}'_{m}\cdots\prod_{m=0}^{n_{\text{jump}}}\sum_{\lambda_{m},\lambda'_{m}}\rho_{\text{foc}}^{(\lambda_{m})}(\mathcal{Z}_{m})\rho_{\text{foc}}^{(\lambda'_{m})}(\mathcal{Z}'_{m}) ,
\end{split}
\end{equation}
where $\mathcal{Z}_{m}$ and $\mathcal{Z}_{m}'$ correspond to the set of Cartesian mapping variables for the forward and backward propagation paths respectively, which are sampled at the beginning of the time interval $t_{m}$ and are then time-evolved over the entire interval. The Cartesian mapping variables for different time intervals are not independent from one another, due to the definition of the real-time correlation function in Eq.~(\ref{eq:corr_jump_PLDM}), which contains matrix products of Stratonovich--Weyl kernels associated with different time intervals. In contrast, the nuclear phase-space variables within the jump spin-PLDM scheme are not re-sampled at intermediate times. Additionally, using focused sampling for the `jumps' within jump spin-PLDM means that for a single trajectory, the electronic state discontinuously hops to the various diabatic states $\lambda$ at certain intermediate times, in close analogy (but not equivalent)\cite{Martens2020} to Tully's fewest-switches surface hopping.\cite{Tully1990hopping}

In order to calculate a real-time correlation function up to time $t$ using $n_{\text{jump}}$ jumps, the jump spin-PLDM method is implemented as follows. First, the time is split up into $n_{\text{jump}}+1$ intervals, each of length $\Delta t=t/(n_{\text{jump}}+1)$. For times $0\leq t'< \Delta t$, the real-time correlation function is calculated using the standard spin-PLDM method, with no jumps (i.e., Eq.~(\ref{eq:corr_jump_PLDM}) with $t_{0}=t'$ and $n_{\text{jump}}=0$). Then for times $\Delta t\leq t'< 2\Delta t$, the correlation function is calculated using one jump within the jump spin-PLDM method, with $t_{0}=\Delta t$ and $t_{1}=t'-\Delta t$. Performed in this way, the jumps are then always applied at the end of every $\Delta t$ time interval (except for the end of the last interval). Practically, the real-time correlation function at all intermediate times can be obtained by averaging over trajectories, each of which is associated with a given instance of the $n_{\text{jump}}$ sets of mapping variables for the forward and backward paths, $\{\mathcal{Z}_{0},\mathcal{Z}_{1},\cdots,\mathcal{Z}_{n_{\text{jump}}}\}$ and $\{\mathcal{Z'}_{0},\mathcal{Z'}_{1},\cdots,\mathcal{Z'}_{n_{\text{jump}}}\}$. While this is the simplest implementation of the jump spin-PLDM scheme, more advanced schemes could be used to improve the convergence of the method.\cite{Hsieh2013FBTS} 

\begin{figure}
\includegraphics{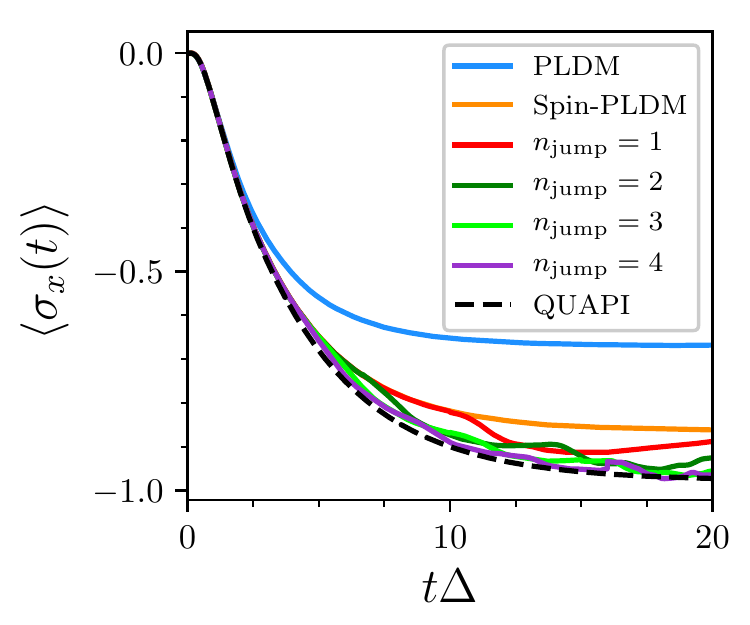}
\caption{Calculation of the $\braket{\sigma_{x}(t)}$ expectation value for jump spin-PLDM, for different numbers of evenly-spaced jumps. The results correspond to a low temperature symmetric Ohmic spin-boson model [Model (b) in Paper I].\cite{paper1} The system is initially prepared in the excited electronic state $\ket{1}$. The dashed black line gives the numerically exact result, obtained using QUAPI and the blue line gives the standard PLDM/FBTS result.}\label{fig:low}
\end{figure}

Results for the jump spin-PLDM approach are given by Fig.~\ref{fig:low}, which corresponds to the $\braket{\sigma_{x}(t)}$ expectation value for the symmetric low temperature spin-boson model considered previously in Paper I.\cite{paper1} The $\braket{\sigma_{x}(t)}$ expectation value for this spin-boson model was chosen, as this was one of the few situations in which spin-PLDM was seen to produce a significant error in the long-time limit. Fig.~\ref{fig:low} shows that as the number of jumps in jump spin-PLDM is increased, the obtained results systematically converge towards the numerically exact QUAPI result. While the $n_{\text{jump}}=4$ result still shows some deviation from the numerically exact result, it is clear on comparison with results for fewer jumps that this error will disappear when more jumps are performed within the scheme. This is to be expected, because the spin-boson model contains a harmonic bath, for which the linearization approximation for the nuclear forward and backward paths applied within the derivation of jump spin-PLDM is exact. The fact that the jump spin-PLDM result is seen to approach convergence to the numerically exact result with increasing number of jumps further illustrates that the focused conditions implemented within jump spin-PLDM are rigorous and do not constitute an additional approximation when used in conjunction with this method. This is in contrast to focused conditions used for MMST mapping-based approaches. Also shown in Fig.~\ref{fig:low} is the standard PLDM/FBTS result, which exhibits a much greater error than spin-PLDM, as well as also deviating from the numerically exact result at much shorter times. This therefore implies that a fewer number of jumps will be needed to converge jump spin-PLDM to the numerically exact result compared to jump FBTS.
\section{Comparison of spin-PLDM with other related approaches\label{sec:comp}}
In spin-LSC, the W-kernels are used to represent the observable operators, $\hat{A}$ and $\hat{B}$, within the definition of the real-time quantum correlation function, while in spin-PLDM, the W-kernels are used to represent the time-ordered propagators for the forward and backward paths. Hence because the spin-LSC and spin-PLDM methods use the W-kernels to represent different objects within the real-time correlation function, the associated underlying expressions for both techniques, given by Eqs.~(\ref{eq:corr_lsc_foc}) and (\ref{eq:corr_pldm_foc}), look quite different. In order to compare the methods to each other, we can re-express the spin-PLDM correlation function in terms of the mapping-variable representations of the underlying operators $\hat{A}$ and $\hat{B}$: 
\begin{equation}
\label{eq:pldm_op}
B_{\text{W}}(\mathcal{Z},\mathcal{Z}')=\tfrac{1}{2}\sum_{\lambda,\lambda'}\braket{\lambda|\hat{B}|\lambda'}\left(Z_{\lambda}^{*}Z'_{\lambda'}-\gamma_{\text{W}}\delta_{\lambda\lambda'}\right) .
\end{equation}
We choose this definition of the spin-PLDM operator representation so that it becomes identical to the standard PLDM operator representation [Eq.~(15) of Ref.~\onlinecite{paper1}] when $\gamma_{\text{W}}=0$ and also identical to the spin-LSC operator representation [Eq.~(\ref{eq:w_func2})] when $\mathcal{Z}=\mathcal{Z}'$.

Here, we only consider correlation functions $C_{AB}(t)$ that satisfy $\text{tr}[\hat{B}]=0$, because any operator can be decomposed into a traceless part and the identity and the correlation function $C_{A\mathcal{I}}(t)$ is trivially time-independent. However, $C_{\mathcal{I}B}(t)$ carries important dynamical information. To make the comparison, the identity-containing correlation functions (with $\hat{A}=\hat{\mathcal{I}}$) and the correlation functions of traceless operators (with $\text{tr}[\hat{A}]=0$) are considered separately. This separation is chosen because the expression for identity-containing correlation functions can often be simplified by imposing $\hat{A}=\hat{\mathcal{I}}$ within the underlying derivation. This has already been successfully achieved for fully linearized MMST mapping, where employing an `identity-trick' was found to significantly increase the accuracy of calculating such correlation functions.\cite{identity,FMO,linearized} Because correlation functions containing traceless operators decay to zero in the long-time limit, calculating identity-containing correlation functions accurately is important for obtaining the correct long-time population relaxation. 

The discussion in this section can be equally applied to either the full-sphere or focused variants of the methods.
\subsection{Identity-containing correlation functions}
From Eq.~(\ref{eq:kernel_cartesian_time}), the time evolution of the spin-PLDM correlation function in general does not just depend on the time-evolved mapping variables, but also on the time-ordered electronic propagator, $\hat{U}(t)$. However, when $\hat{A}=\hat{\mathcal{I}}$, such identity-containing correlation functions can be rewritten equivalently in terms of the evolved mapping variables:
\begin{equation}
\begin{split}
\label{eq:identity_exp}
&\!\!\!\!\!\!\!\!\!\!C_{\mathcal{I}B}(t)=\Braket{\mathcal{I}_{\text{W}}(\mathcal{Z},\mathcal{Z}')B_{\text{W}}(\mathcal{Z}'(t),\mathcal{Z}(t))}_{\text{spin-PLDM}} \\
&\!\!\!\!\!\!\!\!\!\!+\frac{\gamma_{\text{W}}}{2}\Braket{FB_{\text{W}}(\mathcal{Z}'(t),\mathcal{Z}(t))-B_{\text{W}}(\mathcal{Z}(t))-B_{\text{W}}(\mathcal{Z}'(t))}_{\text{spin-PLDM}} ,
\end{split}
\end{equation}
as derived in Appendix~\ref{app:identity_cont}. In this expression, $B_{\text{W}}(\mathcal{Z})$ is the W-function for an electronic operator $\hat{B}$, given by Eq.~(\ref{eq:w_func2}) and $B_{\text{W}}(\mathcal{Z},\mathcal{Z}')$ is the spin-PLDM expression for the same operator, given by Eq.~(\ref{eq:pldm_op}). The term $\mathcal{I}_{\text{W}}(\mathcal{Z},\mathcal{Z}')=\frac{1}{2}(\mathcal{Z}^*\cdot\mathcal{Z}'-R_{\text{W}}^{2})+1$, which is also defined by Eq.~(\ref{eq:pldm_op}), acts as an `overlap factor', between the Cartesian mapping variables for the forward and backward electronic paths ($\mathcal{Z}$ and $\mathcal{Z}'$ respectively).  

There are two noteworthy points of Eq.~(\ref{eq:identity_exp}). First, if we set the zero-point energy parameter, $\gamma_{\text{W}}$, to zero, the expression for the identity-containing correlation function depends entirely on the standard PLDM expressions for the underlying electronic operators, $\mathcal{I}_{\text{m}}(\mathcal{Z},\mathcal{Z}')$ and $B_{\text{m}}(\mathcal{Z}'(t),\mathcal{Z}(t))$ defined in paper I.\cite{paper1} This illustrates that spin-PLDM differs from standard PLDM not just through the initial sampling of the Cartesian mapping variables, which are constrained to the hypersphere, but also through the existence of a zero-point energy parameter. The existence of a zero-point energy parameter results in additional terms being present in the expression for the identity-containing correlation function and results in improved long-time populations, as shown in Paper I.\cite{paper1} 

Second, in the case that $\mathcal{Z}=\mathcal{Z}'$, 
the form of Eq.~(\ref{eq:identity_exp}) becomes similar to the expression for the identity-containing correlation function within the spin-LSC method, given by Eq.~(\ref{eq:lsc_ident}), which has only one set of mapping variables. This follows from: $\mathcal{I}_{\text{W}}(\mathcal{Z},\mathcal{Z})=\mathcal{I}_{\text{W}}(\mathcal{Z})=1$ and $B_{\text{W}}(\mathcal{Z}(t),\mathcal{Z}(t))=B_{\text{W}}(\mathcal{Z}(t))$. This result is perhaps unsurprising, because Eq.~(\ref{eq:lsc_ident}) can be derived as an `identity trick'\cite{identity,FMO} for spin-PLDM, as shown in Appendix~\ref{app:identity}. Hence spin-LSC and spin-PLDM treat the identity-containing correlation functions in a similar way that results in more accurate long-time populations than standard PLDM.
\subsection{Correlation functions of traceless operators}
In general, the expression for the correlation functions of traceless operators within spin-PLDM cannot be easily simplified, although it is still simple to evaluate numerically. This is because the evolution of the zero-point energy parameter within Eq.~(\ref{eq:kernel_cartesian_time}) depends on the time-ordered electronic propagator, which cannot be expressed directly in terms of the time-evolved Cartesian mapping variables. However, the correlation functions of traceless operators within spin-PLDM can be rewritten in a form that enables further analysis:
\begin{equation}
\begin{split}
\label{eq:spinpldm_nonident}
C_{AB}(t)&=\frac{1}{F+1}\Braket{A_{\text{W}}(\mathcal{Z},\mathcal{Z}')B_{\text{W}}(\mathcal{Z}'(t),\mathcal{Z}(t))}_{\text{spin-PLDM}}\\
&\qquad+\Delta C_{AB}(t), 
\end{split}
\end{equation}
as shown in Appendix~\ref{app:non-identity_cont}. In this expression $\Delta C_{AB}(t)$ is defined as the difference between the full spin-PLDM correlation function and the first term. 
This separation is chosen such that $\Delta C_{AB}(0)=0$ in general and $\Delta C_{AB}(t)=0$ in the absence of electron-nuclear coupling. While a convenient and general expression for $\Delta C_{AB}(t)$ is hard to obtain, such an expression is easier to derive if just two-level systems are considered.

For two-level systems, any operator can be written in terms of a spin coherent state and a state orthogonal to it. We therefore define a state, $\ket{\mathcal{\zeta}}$, whose Cartesian mapping variables lie on the same hypersphere as $\ket{\mathcal{Z}}$, but are also orthogonal to it:
\begin{equation}
\label{eq:zeta_orth}
\sum_{\lambda=1}^2 \zeta^{*}_{\lambda}Z_{\lambda}=0 . \\
\end{equation}
It can be shown that $\ket{\mathcal{\zeta}}$ corresponds to the inversion of $\ket{\mathcal{Z}}$ in the spin space, described mathematically as:\footnote{Eq.~(\ref{eq:zeta_orth}) is solved as follows: $\zeta_{1}=Z_{2}^{*}$ and $\zeta_{2}=-Z_{1}^{*}$. Inserting these expressions for $\mathcal{\zeta}$ into the expression for $[\sigma_{j}]_{\text{W}}(\mathcal{\zeta})$ [Eq.~(\ref{eq:w_func2})] directly leads to the result given in Eq.~(\ref{eq:zeta_spin})}
\begin{equation}
\label{eq:zeta_spin}
[\sigma_{j}]_{\text{W}}(\mathcal{\zeta})=-[\sigma_{j}]_{\text{W}}(\mathcal{Z}) ,
\end{equation}
for any $j=(x,y,z)$. This relationship between the $Z_{\lambda}$ and $\zeta_{\lambda}$ mapping variables is shown pictorially in Fig.~\ref{fig:spin2}. Using these two sets of mapping variables, the term $\Delta C_{AB}(t)$ can be rewritten as:
\begin{figure}
\resizebox{0.35\textwidth}{!}{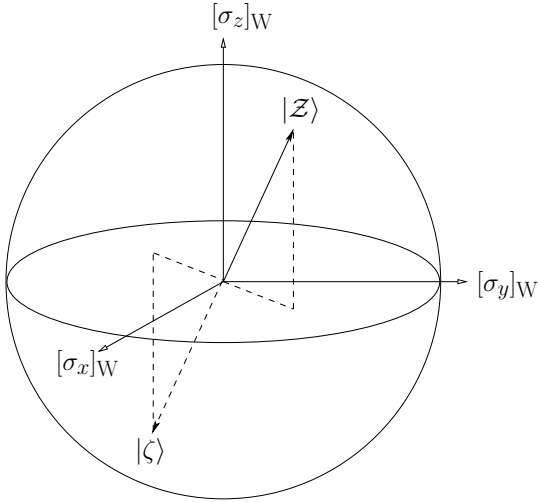}
\caption{An illustration of the $\zeta_{\lambda}$ mapping variables for an $F=2$ system, which lie on the same hypersphere as $Z_{\lambda}$, but are also orthogonal to it. The corresponding $\ket{\mathcal{\zeta}}$ state is the inversion of $\ket{\mathcal{Z}}$ through the origin.}\label{fig:spin2}
\end{figure}
\begin{equation}
\label{eq:spin_correc}
\begin{split}
&\Delta C_{AB}(t)=\frac{\gamma_{\text{W}}}{12}\left(\gamma_{\text{W}}\Braket{A_{\text{W}}(\zeta,\zeta')B_{\text{W}}(\zeta'(t),\zeta(t))}_{\text{spin-PLDM}}\right. \\
&-(\gamma_{\text{W}}+2)\Braket{A_{\text{W}}(\mathcal{Z},\zeta')B_{\text{W}}(\zeta'(t),\mathcal{Z}(t))}_{\text{spin-PLDM}} \\
&-(\gamma_{\text{W}}+2)\Braket{A_{\text{W}}(\zeta,\mathcal{Z}')B_{\text{W}}(\mathcal{Z}'(t),\zeta(t))}_{\text{spin-PLDM}} \\
&\left.+(\gamma_{\text{W}}+4)\Braket{A_{\text{W}}(\mathcal{Z},\mathcal{Z}')B_{\text{W}}(\mathcal{Z}'(t),\mathcal{Z}(t))}_{\text{spin-PLDM}}\right) ,
\end{split}
\end{equation}
where the $\zeta'_{\lambda}$ mapping variables are in the same way orthogonal to $Z'_{\lambda}$. The derivation of this expression is outlined in Appendix~\ref{app:non-identity_cont}. Such a term has never been included within a mapping-based classical-trajectory technique before. For example, Eq.~(\ref{eq:spin_correc}) is zero when $\gamma_{\text{W}}=0$ and hence such a term is not present within standard PLDM\@. In addition, the first term on the right hand side of Eq.~(\ref{eq:spinpldm_nonident}) has a similar form to the full spin-LSC correlation function [Eq.~(\ref{eq:corr_LSC})] when $\mathcal{Z}=\mathcal{Z}'$ and therefore spin-LSC also has no analogue to $\Delta C_{AB}(t)$. This additional $\Delta C_{AB}(t)$ term therefore constitutes the main difference between the spin-LSC and spin-PLDM correlation functions, in addition to the fact that the later contains twice as many Cartesian mapping variables.

The presence of $\mathcal{\zeta}$ and $\mathcal{\zeta}'$ Cartesian mapping variables in Eq.~(\ref{eq:spin_correc}) confirms that this additional term contains effects arising physically from propagating the electronic subsystem on the opposite side of the spin-sphere to that of the coherent state from which the nuclear force is calculated. Because such a term has been rigorously derived within a Stratonovich--Weyl approach to spin-mapping, this suggests that spin-mapping based techniques which only describe the electronic dynamics at a single point of the spin-sphere are in some sense deficient; electronic dynamics at the `antipode' is therefore necessary in order to give a fuller description of the dynamics within coupled electron-nuclear systems. In quantum mechanics, the nuclear degrees of freedom can induce instantaneous excitations of the electronic subsystem and hence the dynamics of the `antipode' do correspond to physical allowed processes within the real system. Hence spin-PLDM is able describe physical processes which are completely neglected in other mapping-based classical-trajectory techniques. 

For systems with an arbitrary number of electronic states, $F$, the term $\Delta C_{AB}(t)$ will involve propagation of a set of $F-1$ states orthogonal to $\mathcal{Z}$ and hence this term physically corresponds to propagating all possible instantaneous excitations of the underlying coherent state. In the next section, the properties and effects of this additional term will be investigated numerically.
\subsection{\label{sec:res}Results}
As in Ref.~\onlinecite{identity}, we present results for $C_{\mathcal{I}\sigma_{z}}(t)$ and $C_{\sigma_{z}\sigma_{z}}(t)$. By taking linear combinations of these two functions, one can recover the more usual measures of state-to-state population transfer. To illustrate the differences between the various methods analysed in this paper, we study the Ohmic spin-boson model used previously in Refs.~\onlinecite{Kelly2016master} and \onlinecite{identity}. This model is both asymmetric and at a relatively low temperature. The parameters used for this spin-boson model also correspond to an intermediate regime between strongly incoherent decay and coherent oscillations, which makes it one of the more challenging spin-boson models to study.
All calculations are performed with $f=100$ nuclear degrees of freedom and $10^{6}$ trajectories. Additionally, numerically exact results are obtained using the quasiadiabatic path-integral (QUAPI) technique.\cite{Makarov1994}  

We begin by noticing that for both the $C_{\mathcal{I}\sigma_{z}}(t)$ and $C_{\sigma_{z}\sigma_{z}}(t)$ correlation functions obtained with our spin-PLDM method, the accuracy of the result is barely affected whether focused of full-sphere initial conditions are used. This therefore illustrates that spin-mapping focused conditions are no less rigorous than the full-sphere version and do not constitute an additional approximation when implemented within spin-PLDM\@. This is similar to what was observed for spin-LSC in Refs.\onlinecite{spinmap} and \onlinecite{multispin}.
For standard PLDM, however, focused conditions do constitute an additional approximation to the original method and hence we only present results for the most accurate non-focused variant. 

\begin{figure*}
\includegraphics[width=\textwidth]{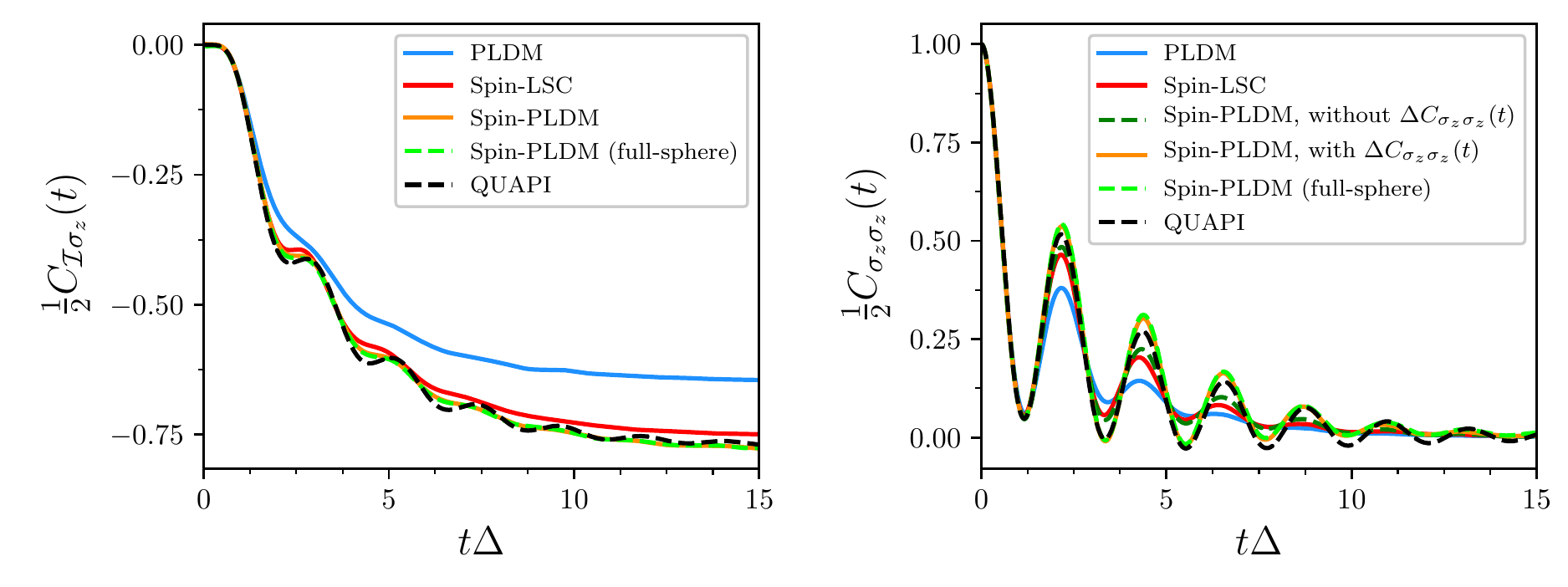}
\caption{Comparison of the $C_{\mathcal{I}\sigma_{z}}$ and $C_{\sigma_{z}\sigma_{z}}$ correlation functions for standard PLDM, focused spin-LSC and spin-PLDM in both full-sphere and (where not otherwise indicated) focused variants. The results correspond to the Ohmic spin-boson model with parameters: $\epsilon=\Delta$, $\xi=0.2$, $\beta\Delta=10$ and $\omega_{\text{c}}=2.5\Delta$. The dashed black lines give the numerically exact results, obtained using QUAPI.}\label{fig:compar}
\end{figure*}

For the identity-containing $C_{\mathcal{I}\sigma_{z}}(t)$ correlation function, both spin-LSC and spin-PLDM methods produce similarly accurate results. This is not surprising, as the expression for the $C_{\mathcal{I}\sigma_{z}}(t)$ correlation functions for both of these methods, given by Eqs.~(\ref{eq:lsc_ident}) and (\ref{eq:identity_exp}), have a similar form. Additionally, the spin-LSC expression (given by Eq.~(\ref{eq:lsc_ident})) can be derived from an `identity trick' applied to spin-PLDM, as shown in Appendix~\ref{app:identity}, which suggests that both methods treat the identity-containing correlation functions on an equal footing. Therefore Fig.~\ref{fig:compar} shows that having two electronic mapping variables in spin-PLDM only results in a small improvement to the spin-LSC $C_{\mathcal{I}\sigma_{z}}(t)$ correlation function for this parameter regime. In contrast, the spin-mapping based method produces significantly more accurate results than standard PLDM, which has the incorrect asymptote at long times. This illustrates a clear advantage of sampling the initial Cartesian mapping variables from the surface of a hypersphere.

For the $C_{\sigma_{z}\sigma_{z}}(t)$ correlation function, the solid green line shows the (focused) spin-PLDM result, given by Eq.~(\ref{eq:spinpldm_nonident}), without including the term $\Delta C_{\sigma_{z}\sigma_{z}}(t)$. Hence the only difference between this result and the $C_{\sigma_{z}\sigma_{z}}(t)$ correlation function for the spin-LSC method is that the spin-PLDM expression contains two electronic mapping variables. As for the identity-containing correlation functions, having two sets of mapping variables only leads to a small improvement in the $C_{\sigma_{z}\sigma_{z}}(t)$ correlation function for this model. The main improvement of spin-PLDM over spin-LSC is hence contained in the term $\Delta C_{\sigma_{z}\sigma_{z}}(t)$, which corrects for the overdamped coherences observed in the correlation functions of traceless operators calculated using spin-LSC\@. Even more severe overdamping is observed in correlation functions of traceless operators calculated using standard PLDM, perhaps due to leakage out of the physical subspace.
\section{Conclusions}
In this paper we have provided a thorough analysis of the spin-PLDM method by identifying the terms present in the real-time correlation function that are responsible for its improved accuracy compared to other mapping-based classical-trajectory techniques. In addition, the spin-PLDM method derived in Paper I\cite{paper1} is made more efficient by the introduction of focused initial conditions, which require fewer trajectories to converge dynamical observables with their associated accuracy seemingly unaffected. The rigorous nature of these focused initial conditions are then demonstrated by using them to obtain a jump spin-PLDM scheme, which is shown to systematically improve spin-PLDM results to those of the QCLE\@. Even though this is computationally intensive approach, such a scheme can at the very least be used practically to quantify the error associated with calculated results,\cite{Kelly2020_2} which is often difficult to do with other mapping-based classical-trajectory techniques.  

Spin-PLDM appears to be able to improve upon previously derived fully linearized approaches such as spin-LSC and also upon the standard PLDM approach based on MMST mapping. In particular, we have shown that spin-PLDM contains a term which is absent in spin-LSC, and which appears to solve the problem of `overdamped coherences' observed in the correlation functions of traceless operators calculated using spin-LSC\@. Such a term is also absent within standard PLDM, which also suffers with the same problem of `overdamped coherences'. However a more significant improvement over standard PLDM by spin-PLDM is observed in the identity-containing correlation functions, where confining the Cartesian mapping variables to the surface of a hypersphere in spin-mapping is shown to lead to a significant improvement in the accuracy of such correlation functions in the long-time limit.

Together with focused inital conditions and the jump scheme, spin-PLDM is shown to be a powerful tool for accurately describing non-adiabatic dynamics in large condensed phase systems. Being based on classical trajectories makes the method relatively computationally cheap, while the ability to systematically improve the accuracy of its results means it also retains some of the advantages of more expensive numerically exact wavefunction based methods. While such a method cannot at the moment describe quantum nuclear effects, there is promise that such effects could be included in the future through a ring-polymer formulation of the theory.\cite{mapping,Ananth2013MVRPMD,Chowdhury2017CSRPMD,Tao2018isomorphic}

Several questions still remain unanswered. First, is there some framework by which the accuracy of mapping-based techniques can be compared, so that some definitive consensus can be made on which techniques are the most reliable and accurate? Second, how close can mapping-based classical-trajectory techniques involving independent trajectories get to an exact solution of the QCLE? In other words, what would be the ultimate mapping-based technique and in what limits would it be able to exactly describe the dynamics of systems?
\begin{acknowledgments}
The authors would like to acknowledge the support from the Swiss National Science Foundation through the NCCR MUST (Molecular Ultrafast Science and Technology) Network. We also thank Johan Runeson for helpful discussions and for his comments made on the original manuscript.
\end{acknowledgments}

\section*{Data Availability}
The data that supports the findings of this study are available within the article.

\appendix

\section{Focused initial conditions}\label{app:foc}
Spin-mapping focused conditions were initially designed to correctly satisfy the Stratonovich--Weyl property given by Eq.~(\ref{eq:sw_foc}) for purely diagonal population operators. In this appendix, we prove that the focused initial conditions introduced in Sections~\ref{sec:foc_lsc} and \ref{sec:foc} satisfy this Stratonovich--Weyl property for any two operators $\hat{A}$ and $\hat{B}$ and can therefore be rigorously used to initially sample the Cartesian mapping variables for evaluating any spin-LSC and spin-PLDM correlation function. While this additional proof is unimportant for spin-LSC when calculating correlation functions involving an initial population operator, this is nevertheless necessary for spin-PLDM, because the W-kernels are used to represent the propagators rather than the observable operators. In particular, we wish to show that:
\begin{equation}
\label{eq:t=0_corr}
\begin{split}
G_{mnn'm'}&=\langle [\ket{m}\bra{n}]_{\text{W}}(\mathcal{Z})[\ket{n'}\bra{m'}]_{\text{W}}(\mathcal{Z})\rangle_{\text{spin-LSC}}^{\text{foc}}\\&=\tr[\ket{m}\braket{n|n'}\bra{m'}]
=\delta_{mm'}\delta_{nn'} ,
\end{split}
\end{equation}
where these states are chosen to be a complete set of orthonormal basis states for the electronic system (i.e, $\braket{m|n}=\delta_{mn}$). Additionally, the W-functions associated with these electronic operators, $[\ket{m}\bra{n}]_{\text{W}}(\mathcal{Z})$, sampled with focused initial conditions can be obtained from Eqs.~(\ref{eq:w_func2}) and (\ref{eq:arctic_param}) as:
\begin{equation}
\label{eq:W-func-op}
[\ket{m}\bra{n}]_{\text{W}}(\mathcal{Z})=\tfrac{1}{2}\left(r_{m}r_{n}\eu{i\left(\phi_{n}-\phi_{m}\right)}-\gamma_{\text{W}}\delta_{mn}\right) .
\end{equation}
where $\phi_{m}$ is an angle uniformly sampled from $0$ to $2\pi$, while $r_{m}$ is given by Eq.~(\ref{eq:r_def}), where $\lambda$ is the diabatic state onto which the Cartesian mapping variables are focused.

Terms in the ensemble average [Eq.~(\ref{eq:t=0_corr})] will only be non-zero if no phase factors arising from W-functions [Eq.~(\ref{eq:W-func-op})] remain.
There are only two ways in which this can occur, either when $m=n$ and $m'=n'$, or when $m=m'\ne n=n'$.
We thus consider the two terms:
\begin{subequations}
\begin{align}
    G_{mmm'm'} &= \tfrac{1}{4}\sum_{\lambda}\left(r_{m}^{2}-\gamma_{\text{W}}\right)\left(r_{m'}^{2}-\gamma_{\text{W}}\right) , \label{eq:case1} \\
    G_{mnnm} &= \tfrac{1}{4}\sum_{\lambda}r^{2}_{m}r^{2}_n,\quad\text{for}\quad m\neq n . \label{eq:case2}
\end{align}
\end{subequations}

In the first case, using Eq.~(\ref{eq:r_def}) leads to the result $G_{mmm'm'}=\delta_{mm'}$. 
In a similar way, the second case [Eq.~(\ref{eq:case2})] can be rewritten as
\begin{equation}
\label{eq:w_prop_coh}
G_{mnnm}=\tfrac{1}{4}\left\{2\gamma_{\text{W}}(\gamma_{\text{W}}+2)+(F-2)\gamma^{2}_{\text{W}}\right\} ,
\end{equation}
when $m\neq n$. The first term on the right hand side of this expression comes from the two terms within the sum over $\lambda$ where the Cartesian mapping variables are focused onto states $m$ and $n$. The second term on the right hand side of Eq.~(\ref{eq:w_prop_coh}) then comes for the remaining $F-2$ terms within this sum over $\lambda$. The result in Eq.~(\ref{eq:w_prop_coh}) can be further simplified by using the expression for the zero-point energy parameter, $\gamma_{\text{W}}$, given by Eqs.~(\ref{eq:gamma_w}) and (\ref{eq:wsphere_radius}), which leads to $G_{mnnm}=1$. All these results are consistent with Eq.~(\ref{eq:t=0_corr}) and this therefore proves that the focused initial conditions on the W-sphere do indeed satisfy the Stratonovich--Weyl properties of the W-functions for any two operators, $\hat{A}$ and $\hat{B}$.
\section{Spin-PLDM identity-containing correlation functions}\label{app:identity_cont}
Within this paper, we show that the expression for the spin-PLDM real-time correlation function can be simplified in the case when $\hat{A}=\hat{\mathcal{I}}$, such that the time-dependence of this correlation function only depends on the time evolution of the Cartesian mapping variables. In order to show this, the following expression for the W-kernel can be used:
\begin{equation}
\label{eq:kernel_cartesian_time_APP}
\hat{w}_{\text{W}}(\mathcal{Z},t)=\tfrac{1}{2}\left(\sum_{\lambda,\mu}Z_{\mu}(t)\ket{\mu}\bra{\lambda}Z_{\lambda}^{*}-\gamma_{\text{W}}\hat{U}(t)\right) ,
\end{equation}
which is simply an analogous expression to Eq.~(\ref{eq:kernel_cartesian_time}). Inserting Eq.~(\ref{eq:kernel_cartesian_time_APP}) into the spin-PLDM identity-containing correlation function [Eq.~(\ref{eq:corr_PLDM}), with $\hat{A}=\hat{\mathcal{I}}$] results in:
\begin{equation}
\begin{split}
\label{eq:identity_exp_APP}
C_{\mathcal{I}B}(t)&=\Braket{\mathcal{I}_{\text{W}}(\mathcal{Z},\mathcal{Z}')B_{\text{W}}(\mathcal{Z}'(t),\mathcal{Z}(t))}_{\text{spin-PLDM}} \\
&+\frac{\gamma_{\text{W}}F}{2}\Braket{B_{\text{W}}(\mathcal{Z}'(t),\mathcal{Z}(t))}_{\text{spin-PLDM}} \\
&-\frac{\gamma_{\text{W}}}{2}\Braket{\tr\left[\hat{U}^{\dagger}(t)\hat{B}\hat{w}_{\text{W}}(\mathcal{Z},t)\right]}_{\text{spin-PLDM}} \\
&-\frac{\gamma_{\text{W}}}{2}\Braket{\tr\left[\hat{w}^{\dagger}_{\text{W}}(\mathcal{Z}',t)\hat{B}\hat{U}(t)\right]}_{\text{spin-PLDM}} \\
&-\frac{\gamma^{2}_{\text{W}}}{4}\Braket{\tr\left[\hat{U}^{\dagger}(t)\hat{B}\hat{U}(t)\right]}_{\text{spin-PLDM}}.
\end{split}
\end{equation}
The first two terms on the right hand side of Eq.~(\ref{eq:identity_exp_APP}) result from replacing both W-kernels in the definition of the spin-PLDM correlation function with the first term in the definition of the W-kernel, given in Eq.~(\ref{eq:kernel_cartesian_time_APP}). The spin-PLDM representation of the operator $\hat{\mathcal{I}}$ in terms of the Cartesian mapping variables, $\mathcal{I}_{\text{W}}(\mathcal{Z},\mathcal{Z}')$, is given by Eq.~(\ref{eq:pldm_op}) and $B_{\text{W}}(\mathcal{Z}',\mathcal{Z})=\tfrac{1}{2}\sum_{\lambda\lambda'}\braket{\lambda|\hat{B}|\lambda'}Z_{\lambda}^{*}Z'_{\lambda'}$ because we have chosen to only consider correlation functions for which $\text{tr}[\hat{B}]=0$. This constitutes no loss of generality because correlation functions of the form $C_{A\mathcal{I}}(t)$ are time-independent and can thus be calculated without requiring a dynamical method.

Additionally, because the trace is invariant to cyclic permutations, the last term on the right hand side of Eq.~(\ref{eq:identity_exp_APP}) is zero, using again that $\text{tr}[\hat{B}]=0$. To simplify the third and fourth term on the right hand side of Eq.~(\ref{eq:identity_exp_APP}), we make use of the following property involving the W-kernel:
\begin{equation}
\label{eq:w_kernel_lin}
\hat{w}_{\text{W}}(\mathcal{Z},t)\hat{U}^{\dagger}(t)=\hat{U}(t)\hat{w}_{\text{W}}(\mathcal{Z})\hat{U}^{\dagger}(t)= \hat{w}_{\text{W}}(\mathcal{Z}(t)),   
\end{equation}
which follows from Eq.~(\ref{eq:kernel_cartesian_time_APP}). Inserting Eq.~(\ref{eq:w_kernel_lin}) into the third and fourth term on the right hand side of Eq.~(\ref{eq:identity_exp_APP}) shows that the time-dependence of $C_{\mathcal{I}B}(t)$ within spin-PLDM can be described entirely by evolving the Cartesian mapping variables, $\mathcal{Z}(t)$ and $\mathcal{Z}'(t)$. Finally, using the definition of the W-function of operator $\hat{B}$ [Eq.~(\ref{eq:w_func})] results in the expression for $C_{\mathcal{I}B}(t)$ given by Eq.~(\ref{eq:identity_exp}).
\subsection{The spin-PLDM `identity trick'}\label{app:identity}
An alternative expression for the identity-containing correlation function can also be obtained by using $\hat{A}=\hat{\mathcal{I}}$ right at the beginning of the derivation of the spin-PLDM correlation function. In Paper I,\cite{paper1} it was shown that applying the linearization approximation to the nuclear degrees of freedom within the path-integral representation of the real-time correlation function, $C_{AB}(t)$, leads to the following expression:  
\begin{equation}
\label{eq:corr_lin}
\begin{split}
&C_{AB}(t)\simeq\sum_{\lambda,\lambda'}\sum_{\mu,\mu'}\braket{\lambda|\hat{A}|\lambda'}\braket{\mu'|\hat{B}|\mu}\int\rd x_{0}\,\rd p_{0}\,\rho_{\text{b}}(x_{0},p_{0}) \\
&\times\int\frac{\rd\Delta p_{0}}{(2\pi)^{f}}\rd x_{N}\prod_{k=1}^{N-1}\rd x_{k}\frac{\rd p_{k}}{(2\pi)^{f}}\rd \Delta x_{k}\frac{\rd\Delta p_{k}}{(2\pi)^{f}}T'_{[\lambda',\mu']}T_{[\mu,\lambda]}\eu{i\Delta S_{0}} .
\end{split}
\end{equation}
In the case where $\hat{A}=\hat{\mathcal{I}}$, the summation over the $\lambda$ and $\lambda'$ indices in Eq.~(\ref{eq:corr_lin}) can be performed explicitly. Hence, the contribution to the real-time quantum correlation function from the electronic transition amplitudes becomes:
\begin{equation}
\label{eq:trans_amp_ident}
\begin{split}
&\sum_{\lambda}T_{[\mu,\lambda]}T'_{[\lambda,\mu']}\simeq \\ &\bra{\mu}\eu{-i\hat{V}(x_{N})\epsilon}\eu{-\tfrac{i}{2}\nabla\hat{V}(x_{N-1})\epsilon \Delta x_{N-1}}\eu{-i\hat{V}(x_{N-1})\epsilon}\cdots \notag\\
&\times\eu{-\tfrac{i}{2}\nabla\hat{V}(x_{1})\epsilon\Delta x_{1}}\eu{-i\hat{V}(x_{1})\epsilon}\eu{+i\hat{V}(x_{1})\epsilon}\eu{-\tfrac{i}{2}\nabla\hat{V}(x_{1})\epsilon\Delta x_{1}}\cdots \\
&\times\eu{+i\hat{V}(x_{N-1})\epsilon}\eu{-\tfrac{i}{2}\nabla\hat{V}(x_{N-1})\epsilon\Delta x_{N-1}}\eu{+i\hat{V}(x_{N})\epsilon}\ket{\mu'} ,
\end{split}
\end{equation}
where we have used $\hat{\mathcal{I}}=\sum_{\lambda}\ket{\lambda}\bra{\lambda}$. This expression can then be approximated in terms of Cartesian mapping variables by inserting $\hat{\mathcal{I}}=\frac{F}{\mathcal{N}}\int\rd\mathcal{Z}_{0}\,\hat{w}_{\text{W}}(\mathcal{Z}_{0})\delta(|\mathcal{Z}_{0}|^{2}-R_{\text{W}}^{2})$ in between the $\eu{\pm i\hat{V}(x_{1})\epsilon}$ operators. Here, $\mathcal{N}$ is a normalization constant given by
$\mathcal{N}=\int\rd\mathcal{Z}_{0}\,\delta(|\mathcal{Z}_{0}|^{2}-R_{\text{W}}^{2})$.
Performing the same steps as in Paper I\cite{paper1} leads to the following expression:
\begin{equation}
\label{eq:trans_kernel_iden}
\sum_{\lambda}T_{[\mu,\lambda]}T'_{[\lambda,\mu']}\approx \frac{F}{\mathcal{N}}\int\rd\mathcal{Z}\braket{\mu|\hat{w}_{\text{W}}(\mathcal{Z}(t))|\mu'}\delta(|\mathcal{Z}|^{2}-R_{\text{W}}^{2})\eu{-iS_{\text{e}}} .
\end{equation}
Now the time-dependence of the identity-containing correlation function is completely described by the time-evolved Cartesian mapping variables, because time-ordered propagators are now positioned either side of the W-kernel, as in Eq.~(\ref{eq:w_kernel_lin}). Additionally, the electronic action is defined as:
\begin{equation}
\label{eq:kernel_elec_action_iden}
S_{\text{e}}=\sum_{k=1}^{N-1}\nabla H_{\text{W}}(\mathcal{Z}(t_{k}),x_{k})\epsilon\Delta x_{k} ,
\end{equation}
where $t_{k}=k\epsilon$ is the time at time-step $k$. Hence evaluating the identity operator explicitly before inserting the Cartesian mapping variables now leads to an expression which contains a single set of Cartesian mapping variables. Following the rest of the spin-PLDM derivation with Eq.~(\ref{eq:trans_kernel_iden}) leads to an expression for the identity-containing correlation function which is identical to that for spin-LSC, given by Eq.~(\ref{eq:lsc_ident}). The fact that the form of the identity-containing correlation function within spin-LSC can be derived from the underlying spin-PLDM equations suggests that both spin-LSC and spin-PLDM will predict the identity-containing correlation function with a similar accuracy. This is indeed in agreement with the spin-boson results presented in Section \ref{sec:res}.

Such an `identity trick' can also be performed for fully linearized MMST mapping.\cite{identity} For example, the same analysis performed in deriving the spin-PLDM `identity trick' presented here can be equally applied to the derivation of the identity-containing correlation function within LSC-IVR (except that an additional linearization approximation within the MMST mapping space must also be performed). This results in the `single unity' method introduced in Ref.~\onlinecite{identity}, which was found to offer a significant increase in accuracy when calculating identity-containing correlation functions with fully linearized MMST mapping based techniques. \cite{identity,FMO,linearized}
\section{Spin-PLDM correlation functions of traceless operators}\label{app:non-identity_cont} 
For the correlation functions of traceless operators within spin-PLDM, the expression cannot in general be written entirely in terms of the time-evolution of the Cartesian mapping variables, $\mathcal{Z}(t)$ and $\mathcal{Z}'(t)$. This is because the time-evolved W-kernel, given by Eq.~(\ref{eq:kernel_cartesian_time_APP}), also contains a contribution from the time-ordered propagator, $\hat{U}(t)$, proportional to the zero-point energy parameter, $\gamma_{\text{W}}$. In order to analyse this contribution from the time-ordered propagator, we must first split the spin-PLDM correlation function into a part whose time-dependence solely comes from the Cartesian mapping variables and an additional term which depends explicitly on $\hat{U}(t)$. However, in numerical simulations it is generally easier to evaluate spin-PLDM correlation functions using Eq.~(\ref{eq:corr_PLDM}), without using this partitioning of terms, especially for systems with $F>2$. 

In the absence of electron-nuclear coupling, the time-dependence of the correlation functions of traceless operators can, however, be given solely in terms of the time-evolution of the Cartesian mapping variables:\cite{Lucke1999} 
\begin{equation}
\label{eq:corr_lin_spin}
C_{AB}(t)\approx\int\rd x\,\rd p\,\rd\Omega\,\rd\Omega'\,\rho_{\text{b}}(x,p)\braket{\Omega|\hat{A}|\Omega'} 
\braket{\Omega'(t)|\hat{B}|\Omega(t)} .
\end{equation}
This is because PLDM, derived using a spin coherent state basis without a zero-point energy parameter, is also exact in this case. Eq.~(\ref{eq:corr_lin_spin}) is also always exact at $t=0$ for any value of the electron-nuclear coupling. In this expression, $\ket{\Omega}=\sum_{\lambda=1}^F c_{\lambda}\ket{\lambda}$ is a spin coherent state, defined in terms of the amplitudes $c_{\lambda}$. In addition, $\rd\Omega$ is the integration element, defined as:
\begin{equation}
\label{eq:omega_int}
\rd\Omega=F\frac{\prod_{\lambda}\rd c_{\lambda}\,\delta(\sum_{\mu}|c_{\mu}|^{2}-1)}{\int\prod_{\lambda}\rd c_{\lambda}\,\delta(\sum_{\mu}|c_{\mu}|^{2}-1)}  
\end{equation}
and $\rd c_{\lambda}=\rd\text{Re}[c_{\lambda}]\rd\text{Im}[c_{\lambda}]$. In this expression, the factor of $F$ appears such that the spin coherent state integrals satisfy $\tr[\hat{\mathcal{I}}]=\int\rd\Omega=F$. Physically, the Dirac delta function in Eq.~(\ref{eq:omega_int}) guarantees that the coherent state $\ket{\Omega}$ is correctly normalized.

Eq.~(\ref{eq:corr_lin_spin}) can be rewritten in terms of Cartesian mapping variables by using the relation: $c_{\lambda}=Z_{\lambda}/R_{\text{W}}$.\cite{spinmap} This results in:
\begin{equation}
\label{eq:corr_lin_spin_cart}
C_{AB}(t)\approx\frac{1}{F+1}\Braket{A_{\text{W}}(\mathcal{Z},\mathcal{Z}')B_{\text{W}}(\mathcal{Z}'(t),\mathcal{Z}(t))}_{\text{spin-PLDM}} , 
\end{equation}
where we have additionally used the expression for the W-sphere radius given by Eq.~(\ref{eq:wsphere_radius}). The spin-PLDM representation of the operator $\hat{B}$ in terms of the Cartesian mapping variables, $B_{\text{W}}(\mathcal{Z},\mathcal{Z}')$, is given by Eq.~(\ref{eq:pldm_op}) and we have also made use of $\text{tr}[\hat{A}]=\text{tr}[\hat{B}]=0$. Eq.~(\ref{eq:corr_lin_spin_cart}) is the same as the first term in the relation for the correlation functions of traceless operators, given by Eq.~(\ref{eq:spinpldm_nonident}). Eq.~(\ref{eq:corr_lin_spin_cart}) is simpler compared to the spin-PLDM correlation function, because it does not explicitly contain the time-ordered propagator, $\hat{U}(t)$.

When $t\neq0$ or for systems with electron-nuclear coupling, Eq.~(\ref{eq:corr_lin_spin_cart}) is no longer equivalent to the spin-PLDM correlation function and therefore constitutes an additional approximation to it. The difference between Eq.~(\ref{eq:corr_lin_spin_cart}) and the spin-PLDM correlation function given by Eq.~(\ref{eq:corr_PLDM}) can be incorporated into a term, $\Delta C_{AB}(t)$, which generally must be calculated in order to accurately obtain correlation functions of traceless operators. By definition this term is hence given by:
\begin{equation}
\begin{split}
\label{eq:deltaC_def}
\Delta C_{AB}(t)&=\Braket{\tr\left[\hat{A}\hat{w}^{\dagger}_{\text{W}}(\mathcal{Z}',t)\hat{B}\hat{w}_{\text{W}}(\mathcal{Z},t)\right]}_{\text{spin-PLDM}} \\
&-\frac{1}{F+1}\Braket{A_{\text{W}}(\mathcal{Z},\mathcal{Z}')B_{\text{W}}(\mathcal{Z}'(t),\mathcal{Z}(t))}_{\text{spin-PLDM}} ,
\end{split}
\end{equation}
which through the definition of the W-kernel in Eq.~(\ref{eq:kernel_cartesian_time_APP}) depends on the time-ordered propagator, $\hat{U}(t)$.

For the case of $F=2$, Eq.~(\ref{eq:deltaC_def}) can be further simplified. First, the electronic identity operator can be expressed in terms of Cartesian mapping variables:
\begin{equation}
\label{eq:identity}
\hat{\mathcal{I}}=\frac{1}{R^{2}_{\text{W}}}\left(\sum_{\lambda,\mu}Z_{\mu}\ket{\mu}\bra{\lambda}Z_{\lambda}^{*} + \sum_{\lambda,\mu}\zeta_{\mu}\ket{\mu}\bra{\lambda}\zeta_{\lambda}^{*}\right) . 
\end{equation}
The first term in this expression corresponds to the spin-coherent state outer product, $\ket{\Omega}\bra{\Omega}$, as can be seen from the relation $c_{\lambda}=Z_{\lambda}/R_{\text{W}}$. The second term contains new Cartesian mapping variables, $\zeta_{\lambda}$, which as defined in Eq.~(\ref{eq:zeta_orth}) are orthogonal to $Z_{\lambda}$. Defined in this way, the $\zeta_{\lambda}$ mapping variables correspond to the electronic state on the opposite side of the spin-sphere to $\ket{\Omega}$, as illustrated in Fig.~\ref{fig:spin2}. Hence this second term is just the outer product of an electronic state orthogonal to the coherent state, $\ket{\Omega}$. Applying the time-evolved propagator, $\hat{U}(t)$, to the left of Eq.~(\ref{eq:identity}) leads to:
\begin{equation}
\label{eq:2-state-propagator}
\hat{U}(t)=\frac{1}{R^{2}_{\text{W}}}\left(\sum_{\lambda,\mu}Z_{\mu}(t)\ket{\mu}\bra{\lambda}Z_{\lambda}^{*} + \sum_{\lambda,\mu}\zeta_{\mu}(t)\ket{\mu}\bra{\lambda}\zeta_{\lambda}^{*}\right) . 
\end{equation}
Inserting this expression for the time-evolved propagator into the expression for the time-evolved W-kernel, given by Eq.~(\ref{eq:kernel_cartesian_time_APP}), leads to a new expression for the W-kernel solely in terms of Cartesian mapping variables:
\begin{equation}
\label{eq:2-state-kernel}
\begin{split}
\hat{w}_{\text{W}}(\mathcal{Z},t)=\frac{1}{2R^{2}_{\text{W}}}\left((\gamma_{\text{W}}+2)\sum_{\lambda,\mu}Z_{\mu}(t)\ket{\mu}\bra{\lambda}Z_{\lambda}^{*}\right. \\
\left.-\gamma_{\text{W}}\sum_{\lambda,\mu}\zeta_{\mu}(t)\ket{\mu}\bra{\lambda}\zeta_{\lambda}^{*}\right) ,
\end{split}
\end{equation}
where we have additionally used $R^{2}_{\text{W}}=2\gamma_{\text{W}}+2$, which is valid when $F=2$. Inserting Eq.~(\ref{eq:2-state-kernel}) into the definition of $\Delta C_{AB}(t)$, given by Eq.~(\ref{eq:deltaC_def}) and expanding the terms leads to the expression for this term given in Eq.~(\ref{eq:spin_correc}). To do this, we note that $R_{W}^{4}=12$ for $F=2$ [Eq.~(\ref{eq:wsphere_radius})], and use the expression for the spin-PLDM representation of the operator $\hat{B}$ in terms of the Cartesian mapping variables, $B_{\text{W}}(\mathcal{Z},\mathcal{Z}')$, given by Eq.~(\ref{eq:pldm_op}). We have again assumed $\text{tr}[\hat{A}]=\text{tr}[\hat{B}]=0$.

\bibliography{paper2}

\end{document}

%% file: spin2.pdf_t
\begin{picture}(0,0)%
\includegraphics{spin2.pdf}%
\end{picture}%
\setlength{\unitlength}{4144sp}%
\begingroup\makeatletter\ifx\SetFigFont\undefined%
\gdef\SetFigFont#1#2#3#4#5{%
  \reset@font\fontsize{#1}{#2pt}%
  \fontfamily{#3}\fontseries{#4}\fontshape{#5}%
  \selectfont}%
\fi\endgroup%
\begin{picture}(7050,7223)(2551,-7350)
\put(9586,-4246){\makebox(0,0)[lb]{\smash{{\SetFigFont{29}{34.8}{\rmdefault}{\mddefault}{\updefault}{\color[rgb]{0,0,0}$[\sigma_{y}]_{\text{W}}$}%
}}}}
\put(5671,-286){\makebox(0,0)[lb]{\smash{{\SetFigFont{29}{34.8}{\rmdefault}{\mddefault}{\updefault}{\color[rgb]{0,0,0}$[\sigma_{z}]_{\text{W}}$}%
}}}}
\put(3286,-5416){\makebox(0,0)[lb]{\smash{{\SetFigFont{29}{34.8}{\rmdefault}{\mddefault}{\updefault}{\color[rgb]{0,0,0}$[\sigma_{x}]_{\text{W}}$}%
}}}}
\put(5986,-3616){\makebox(0,0)[lb]{\smash{{\SetFigFont{29}{34.8}{\rmdefault}{\mddefault}{\updefault}{\color[rgb]{0,0,0}$\theta_{\text{c}}$}%
}}}}
\put(6526,-3886){\makebox(0,0)[lb]{\smash{{\SetFigFont{29}{34.8}{\rmdefault}{\mddefault}{\updefault}{\color[rgb]{0,0,0}$r=\frac{R^{2}_{\text{W}}}{2}$}%
}}}}
\put(3556,-6001){\makebox(0,0)[lb]{\smash{{\SetFigFont{29}{34.8}{\rmdefault}{\mddefault}{\updefault}{\color[rgb]{0,0,0}$[\sigma_{z}]_{\text{W}}(\mathcal{Z})=-1$}%
}}}}
\put(3601,-2401){\makebox(0,0)[lb]{\smash{{\SetFigFont{29}{34.8}{\rmdefault}{\mddefault}{\updefault}{\color[rgb]{0,0,0}$[\sigma_{z}]_{\text{W}}(\mathcal{Z})=1$}%
}}}}
\put(2566,-6631){\makebox(0,0)[lb]{\smash{{\SetFigFont{29}{34.8}{\rmdefault}{\mddefault}{\updefault}{\color[rgb]{0,0,0}$\gamma_{\text{W}}$}%
}}}}
\put(2566,-1614){\makebox(0,0)[lb]{\smash{{\SetFigFont{29}{34.8}{\rmdefault}{\mddefault}{\updefault}{\color[rgb]{0,0,0}$\gamma_{\text{W}}$}%
}}}}
\put(5401,-4696){\makebox(0,0)[lb]{\smash{{\SetFigFont{29}{34.8}{\rmdefault}{\mddefault}{\updefault}{\color[rgb]{0,0,0}$(\phi_{2}-\phi_{1})$}%
}}}}
\end{picture}%

%% file: spin3.pdf_t
\begin{picture}(0,0)%
\includegraphics{spin3.pdf}%
\end{picture}%
\setlength{\unitlength}{4144sp}%
\begingroup\makeatletter\ifx\SetFigFont\undefined%
\gdef\SetFigFont#1#2#3#4#5{%
  \reset@font\fontsize{#1}{#2pt}%
  \fontfamily{#3}\fontseries{#4}\fontshape{#5}%
  \selectfont}%
\fi\endgroup%
\begin{picture}(6967,7201)(2634,-7328)
\put(5671,-286){\makebox(0,0)[lb]{\smash{{\SetFigFont{29}{34.8}{\rmdefault}{\mddefault}{\updefault}{\color[rgb]{0,0,0}$[\sigma_{z}]_{\text{W}}$}%
}}}}
\put(9586,-4246){\makebox(0,0)[lb]{\smash{{\SetFigFont{29}{34.8}{\rmdefault}{\mddefault}{\updefault}{\color[rgb]{0,0,0}$[\sigma_{y}]_{\text{W}}$}%
}}}}
\put(3376,-5416){\makebox(0,0)[lb]{\smash{{\SetFigFont{29}{34.8}{\rmdefault}{\mddefault}{\updefault}{\color[rgb]{0,0,0}$[\sigma_{x}]_{\text{W}}$}%
}}}}
\put(6706,-1681){\makebox(0,0)[lb]{\smash{{\SetFigFont{29}{34.8}{\rmdefault}{\mddefault}{\updefault}{\color[rgb]{0,0,0}$\ket{\mathcal{Z}}$}%
}}}}
\put(4546,-6721){\makebox(0,0)[lb]{\smash{{\SetFigFont{29}{34.8}{\rmdefault}{\mddefault}{\updefault}{\color[rgb]{0,0,0}$\ket{\mathcal{\zeta}}$}%
}}}}
\end{picture}%